# On the Nature of High Field Charge Transport in Reinforced Silicone Dielectrics: Experiment and Simulation

Yanhui Huang(黄彦辉)[*], Linda. S. Schadler

*Department of Material Science and Engineering, Rensselaer Polytechnic Institute, 110 8th street, Troy, New York, 12180, USA*

[*]Correspondence Email: huangy12@rpi.edu

## Abstract

The high field charge injection and transport properties in reinforced silicone dielectrics were investigated by measuring the time-dependent space charge distribution and the current under dc conditions up to the breakdown field, and were compared with properties of other dielectric polymers. It is argued that the energy and spatial distribution of localized electronic states are crucial to determining these properties for polymer dielectrics. Tunneling to localized states likely dominates the charge injection process. A transient transport regime arises due to the relaxation of charge carriers into deep traps at the energy band tails, and is successfully verified by a Monte Carlo simulation using the multiple-hopping model. The charge carrier mobility is found to be highly heterogeneous due to non-uniform trapping. The slow moving electron packet exhibits a negative field dependent drift velocity possibly due to the spatial disorder of traps.

## I. INTRODUCTION

The high field charge transport properties in dielectric polymers and their filled composites are of particular interest as charge transport is strongly associated with energy dissipation, electrical ageing and dielectric breakdown. Many theoretical models have been developed to describe the conduction process, including: the ionic conduction model[1], the Poole-Frenkel model due to thermionic emission from donor states in the bulk[2], the space-charge-limited-current (SCLC) model with or without traps[1], the hopping conduction model with a single[3] or a distribution of trap energies[4]. The latter two models assume that electronic charge carriers are injected from the electrode and involve the concept of "trap", referring to the localized electronic states in the polymer. It has been argued that added fillers can provide additional trap states that can either reduce or enhance the carrier mobility depending on the trap depth[5]. Though different in the elementary process of charge transport, these models all successfully predict the super-linear current-voltage (*I-V*) relationship in the high field regime that has been observed in many polymer systems with or without fillers[6-8]. Distinguishing between these models can be difficult and a unified understanding is still lacking.

The aim of this work is to investigate the high field conduction behavior in $SiO_2$-reinforced silicone rubber, in hopes of complementing the fundamental understanding of charge transport in dielectric polymers and their composites. Silicone rubber is an important dielectric polymer widely used for electronic encapsulation, outdoor insulation and electroactive polymer actuators. Pure silicone rubbers suffer from poor mechanical and dielectric properties. Thus most commercially available silicones are reinforced with a large amount of fumed silica (up to 40 wt%) to improve the mechanical and dielectric properties as well as to lower the cost[9]. Despite its technical importance, the charge transport properties in this type of silicone-based composite are rarely reported[9-12]. In this work, the high field charge transport of $SiO_2$ reinforced silicone rubber was comprehensively studied using pulsed electro-acoustic (PEA) space charge measurements and bulk dc current measurements in a range of fields until dielectric breakdown.

The paper starts with a brief review of the current understanding of the electronic structure of disordered dielectric polymers and the corresponding elementary charge transport mechanisms. Then the charge injection and bulk transport properties in reinforced silicone rubber are presented and thoroughly discussed in comparison with published results on other polymer systems, with a particular focus on rapid charge injection, transient transport and the negative field dependent mobility. These observations offer an opportunity to re-examine the existing models and to improve the fundamental understanding of charge injection and transport in dielectric polymers. It will be shown that the multiple-hopping (MH) model can reasonably describe these phenomena.

## II. MICROSCOPIC UNDERSTANDING OF CHARGE TRANSPORT PROCESS

It has been pointed out that ions are unlikely to make a noticeable contribution to the electrical conduction in dielectric polymers and the dominant charge carrier species are mostly electron/holes that are extrinsically injected[13]. As background, the current understanding of the elementary process of electronic charge carrier transport in dielectric polymers is briefly reviewed.

### A. Electronic charge transport in polymer dielectrics

An organic molecular solid is composed of covalently bonded molecules that are held together by weak van der Waals forces. Dielectric polymers fall into this category and are usually amorphous or semi-crystalline. When a solid looses its crystallographic order, the variation of atomic potential creates localized electronic states at the band tails called Anderson localization[14]. Other factors like impurities and chemical defects also introduce low lying localized states extending into the gap[15-17]. The term "trap" often refers to these localized states. For small degree of disorder, delocalized states may still exist above the localized states, and the energy separating them is known as the mobility edge[14]. There have been many efforts to calculate the mobility edge for dielectric polymers via *ab initio* methods and charge mobilities are calculated assuming band transport[18-20]. On the other hand, extensive investigations in organic semiconductors have revealed that the electron-phonon coupling in molecular solids is very strong and electron transport can lose coherence even in the extended states[21,22]. For incoherent or hopping transport, charge can take the form of phonon-assisted tunneling between adjacent localized states[20-22], and this has been successfully described by a multiple-hopping (MH) model[23]. Excess electrons injected from the cathode travel via localized states in the conduction band while a valence electron extracted at the anode forms a hole that travels in the valence band. For bare electron/hole hopping, the hopping rate can be described using the Miller-Abrahams formula[24],

$$v_{ij} = v_0 \exp(-2\gamma R_{ij}) \times \begin{cases} \exp(-E_{ij}/kT), & E_{ij} > 0 \\ 1, & E_{ij} < 0 \end{cases} \quad (1)$$

wherein the rate for hopping up and down in energy differs by a Boltzmann factor. Here $v_{ij}$ is the hopping rate between occupied site *i* and unoccupied site *j* separated by energy $E_{ij}$ and a distance $R_{ij}$; $v_0$ is the attempt-to-hop frequency on the order of $10^{12}$ Hz and $\gamma^{-1}$ is the decay length of the electronic wave function. The field can reduce the energy barrier by $-R_{ij}Fe$, with *F* being the electric field and *e* being the electron charge. The charge can either jump down to states with equal or lower energies or jump up to states of higher energies. The average hopping rate is closely related to the shape of the density of states (DOS) as a function the energy. At shallow states, the charge will preferentially take a downward jump. But normally the DOS decreases with decreasing energy, so that as the charge relaxes, the density of available states that the charge can jump down to decreases, manifesting as an increasing $R_{ij}$ in the first term in (1) and rendering it more difficult for charges to make downward jumps. At a certain energy point, the rate of downward jumps will become equal to that of upward jumps and this energy level is defined as the transport energy, $E_t$ [23]. Charge carriers at states below $E_t$ will move primarily by jumping upwards to $E_t$ while carriers at states above $E_t$ will move primarily by jumping downwards, so that $E_t$ actually separates localized states that involve mostly downward and upward jumps respectively. The concept of $E_t$ is important in terms of the hopping injection as it characterizes the most probable energy level that the charge jumps to from the electrode[25]. It should be noted that $E_t$ is not the lowest energy that a charge carrier can reach and it will eventually relax into the states at band tails characterized by low density and low energy and it is these deep states that critically impact the steady state mobility.

### B. Charge Transfer at the Metal and Dielectric Interface

Contact electrification experiments show that for many polymer dielectrics like polyethylene (PE), polytetrafluoroethylene (PTFE) and silicone[26-29], or even inorganic dielectrics like $SiO_2$[30] and $Al_2O_3$[31], charge transfer from metal to dielectrics can readily occur upon contact with or without external potential[29,30,32], and that some of the transferred charge can be retained in the dielectric upon breaking the contact. Recent progresses in measuring chemical redox reactions on charged polymer surfaces showed that the charge carrier involved in the exchange for non-ionic polymers has an electronic nature[33]. This is very similar to the charge transfer between metals with different work functions for which the contact is followed by the charge flow and the build up of the contact potential except that for dielectrics the transferred charge are kinetically trapped and can be retained in the dielectric upon separation. Especially, it was found that when a bias is applied to the metal contact, the quantity of transferred charge to the dielectric at equilibrium scales linearly with the external field with a slope of the dielectric capacitance and leaves the contact potential unchanged[29,30,32].

And more surprisingly, the charge transfer is found to be a rather rapid process[29-31], which seems unlikely given the large injection barrier of several eV predicted from the Schottky-Mott limit for wide band gap dielectrics. It is now known that the energy barrier can deviate from the Schottky-Mott limit due to the formation of interface dipoles as a result of complex electronic interaction between two materials[34-36]. But for dielectrics, electrons are strongly bound to the nucleus and the offset is usually small[35]. For instance, Chen *et al.* examined the interface dipole effect at a PE/Al interface using DFT calculations, and found that the interface dipole only causes a small energy shift of 0.2 ~ 0.3 eV, which is still too small to account for the large discrepancy between the experimentally measured energy barrier and the one predicted by theory[37]. Therefore it was often argued that charges can be directly injected to the deep trap states by phonon-assisted tunneling characterized with a much smaller barrier and without necessarily being excited over a larger energy barrier up to the conduction band edge or mobility edge[31,38,39]. This is consistent with the MH model that charge can move directly between different localized states by tunneling. The injected charges do not only dwell on the surface, but can also travel into the bulk and contribute to a continuous current flow, and this has been verified by PEA measurements[40].

## III.   EXPERIMENT AND RESULTS

### A.   Sample preparation

Sylgard® 184 was purchased from Dow Corning to prepare silicone samples. To the best of our knowledge, the silicone resin contains 40 wt% fumed silica. The resin was mixed with hardener in a 10:1 mass ratio using a vortex mixer at 3500 rpm before casting into a flat Al pan and then was cured at 100 $^0$C under vacuum for 1.5 hour. The unfilled silicone was purchased from Gelest with resin as DMS-V25 and hardener as HMS-301, catalyzed by Pt-divinyltetramethyldisiloxane from Sigma-Aldrich. The samples had a thickness of 370 ± 20 μm and 50 ± 5 μm for the PEA and current measurement respectively. For each test, multiple samples (2 for PEA and 5 for current measurements) were used to establish the reproducibility.

### B.   Space charge measurement

The mechanism and setup for the PEA measurement can be found elsewhere[41]. The equipment has an Al bottom electrode as ground and carbon black loaded semi-conductive (SC) polyethylene as the top HVDC electrode. A high voltage source with different polarities was used to study the effect of the electrode. The probe pulse had a width of 10 ns, a repetition frequency of 140 Hz, and an amplitude of 300 V. The signal was obtained by averaging over 512 pulse applications, corresponding to a time span of 3.6 s, and was recorded every 5 seconds. The final charge distribution was obtained by a deconvolution of the raw data.

The space charge evolution at a nominal field of 10 MV/m during and after electric stress is shown in Fig.1a and Fig.1b respectively. The space charge evolution at higher fields was also examined and is plotted in Fig.2. The effect of electrode material on injection was unveiled by flipping the polarity of the voltage and the result is plotted in Fig.1d, which shows a drastically different picture.

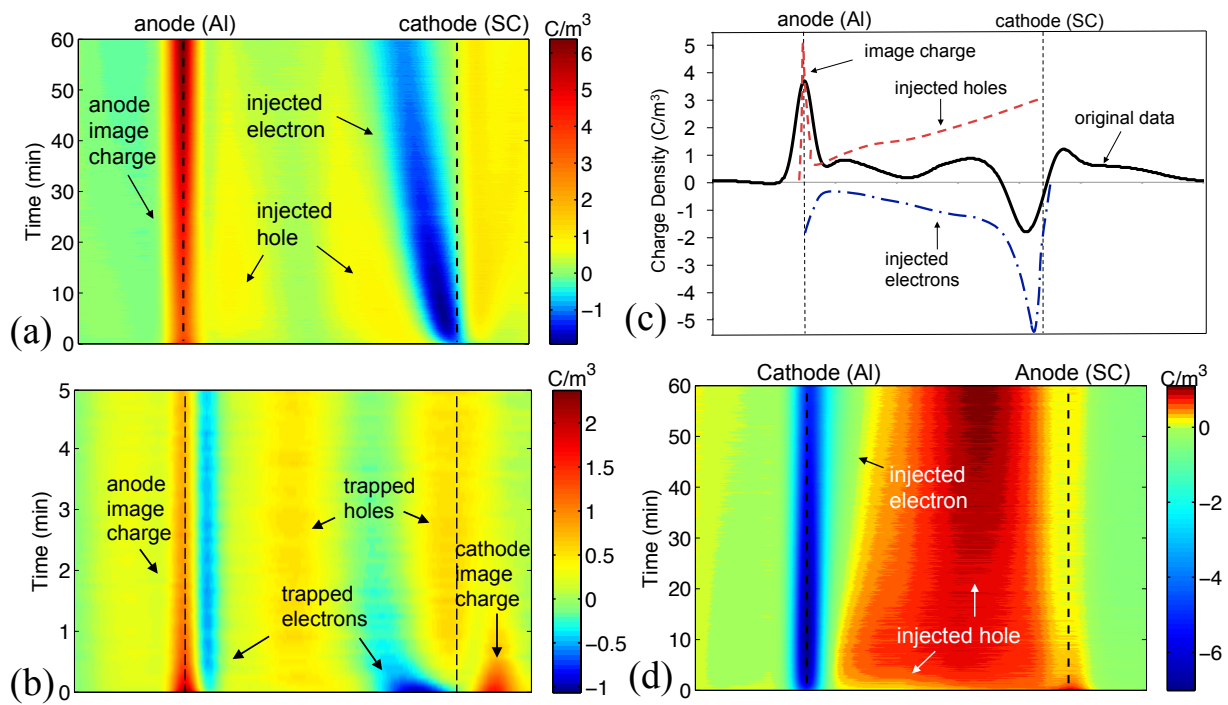

FIG. 1. (a) The space charge evolution in silicone at a nominal field of 10 MV/m; (b) the charge decay profile after 1 hour stressing with voltage off; (c) the estimated bipolar charge distribution in the silicone sample after 10 min stressing; (d) the space charge evolution in silicone under a opposite polarity at -10 MV/m. The x-axis represents the spatial position inside the sample and the dashed line marks the position of the electrode. Due to the large acoustic attenuation of the soft silicone sample, the height of the signal near the top electrode (always plotted on the right) was reduced by approximately 4/5 of its counterpart near the bottom electrode and also broadened.

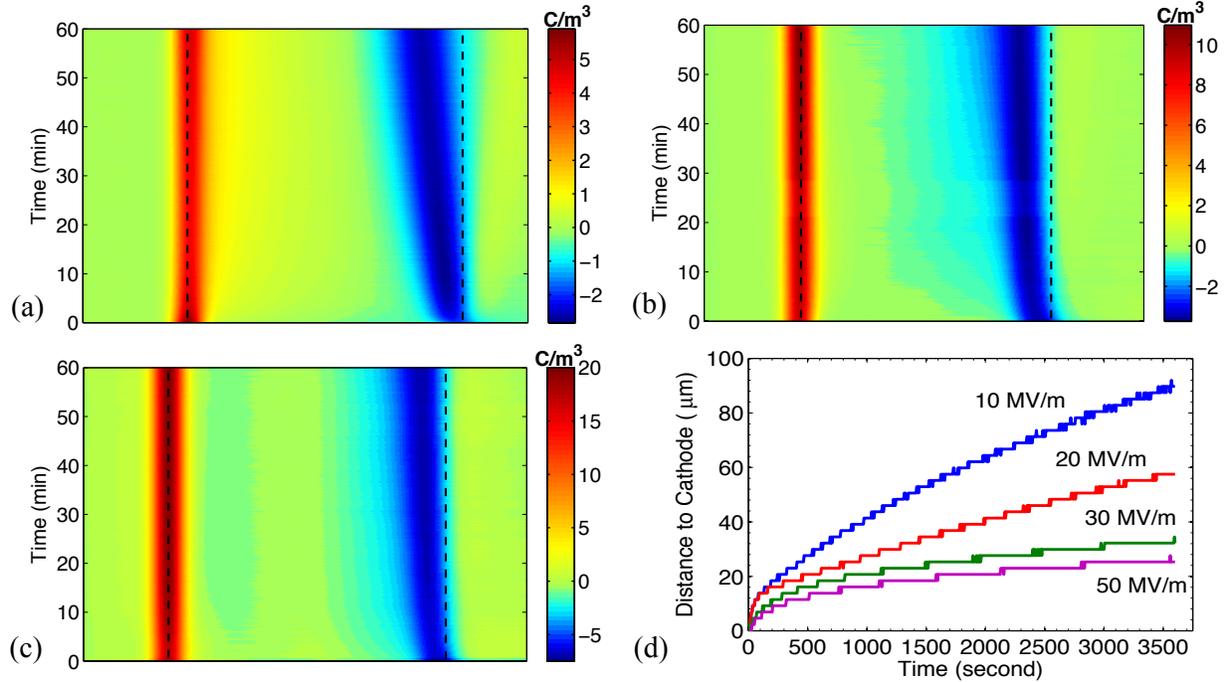

FIG. 2. (a) to (c) The space charge distribution evolution at a nominal field of 20, 30, 50 MV/m respectively; the dashed line marks the position of the electrode (Al anode on the left and SC cathode on the right); (d) The peak location of the electron packet with respect to the cathode as a function of polarization time at different fields.

Bipolar injection and transport was observed, but the spatial charge distribution appears to be different for electrons and holes. As shown in Fig.1a, electrons were quickly injected and formed a packet moving towards the anode. Fig.2d plots the distance of the electron packet peak to the cathode as a function of stressing time, from which we can identify a short transient transport period at the beginning of the stressing, in which the electron packet velocity decreased with time until it reached quasi-steady state after ~100 seconds. Fig.2d also reveals that the velocity of the electron packet decreases with the field, manifesting a negative field dependent mobility. In addition, as shown in Fig.1a and Fig.2, electron transport is found to be dispersive as the width of the electron packet increased with time. Electron accumulation against the anode was observed after stress was applied for some time. This can be first inferred from the observation that the hole region near the anode gradually became disconnected from the anode charge peak (Fig.1a) and was further confirmed from a small negative peak near the anode that is visible after the voltage is removed (Fig.1b). This build up of charge indicates that some electrons have higher mobilities than the charge packet and their extraction rate at the opposite electrode is smaller than the hopping rate in the bulk.

Holes, on the other hand, as shown in Fig.1a and Fig.1d, were more uniformly distributed and did not form a charge packet, exhibiting more dispersive transport. From Fig.1d, hole accumulation at the cathode occurs almost immediately after applying the voltage, implying that the hole extraction rate the SC electrode is also smaller than the hopping rate. Note that this is not obvious at higher fields as it is overwhelmed by injected electrons, but it can still be observed after most injected electrons are depleted after removing the voltage. The charge decay profile field at higher field can be found in the Fig. S1 in the supplemental material[42].

Based on the analysis above, we hypothesize the actual bipolar charge distribution in the sample to match the net charge distribution measured from PEA shown in Fig.1c and the results are sketched in Fig.1c. The acoustic attenuation is left unchanged. It is argued that the real electron density near the cathode should be higher than the PEA measured value due to the existence of holes. The hole concentration on the other hand increases in a slope towards the cathode as result of diffusion. The small valley of positive charges near the anode is due to the accumulation of electrons against the electrode.

### C. Conductivity measurement

The sample tested for dc conduction used the aluminum anode that was used to cast the sample, and a brass cathode with a guarded ring to eliminate the surface current. To study the effect of the electrode on conductivity, a semi-conductive polyethylene tape and a 17 μm biaxially-oriented polypropylene (PP) dielectric film were inserted respectively between the sample and the guarded brass cathode. As shown in Fig.3, the change in the output current is minimal and within the magnitude of test variance.

Thinner samples of 50 μm thickness were used to measure the conducting current at higher field. The voltage was increased in a stepwise manner of 500 V increments, and 10 min was allowed for stabilization at each voltage until dielectric breakdown. The current profile is shown in Fig.4. The current increases in an almost linear fashion with the field with current spikes showing up at fields close to breakdown. And more interestingly, as shown in Fig.5, while the electron packet velocity $u_p$ decreases with the field, the measured current still increases with the field.

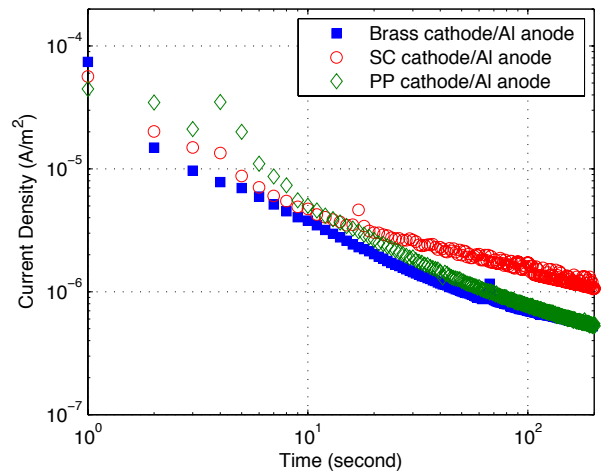

FIG. 3. The current density $J_0$ measured in the external circuit using different cathode material at 10 MV/m. Thickness of the sample used here is 400 um to make it comparable to the results of PEA.

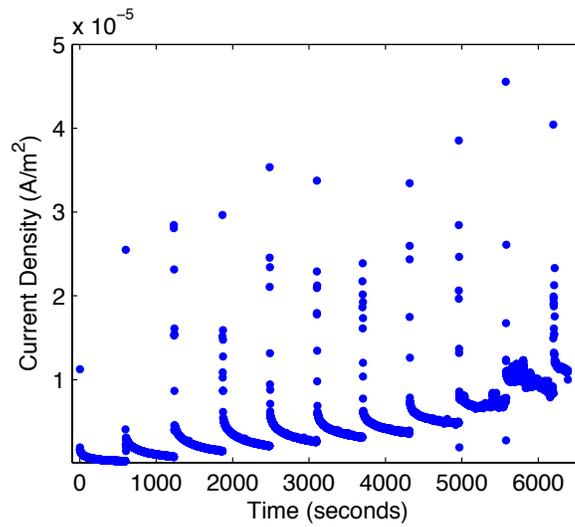

FIG. 4. The current profile measured by ramping the field stepwise every 5 MV/m for 600 seconds until breakdown.

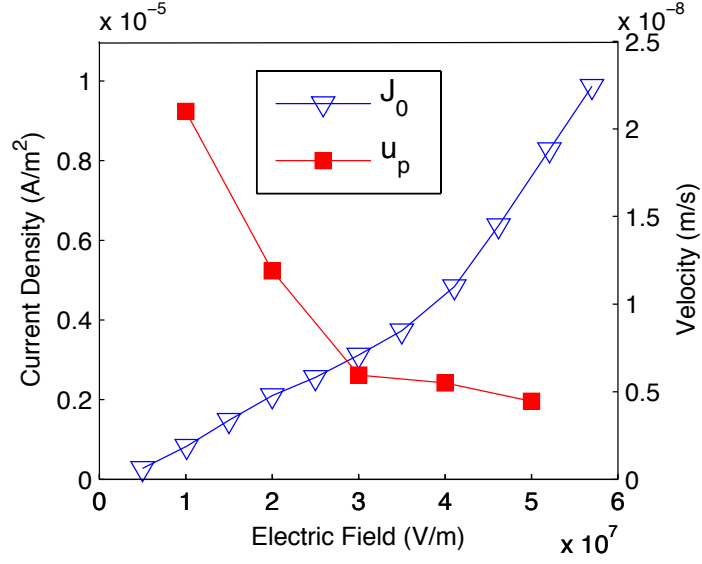

FIG. 5. The measured quasi-steady state current density $J_0$, electron packet moving velocity $u_p$ as a function of field.

## IV. DISCUSSION

### A. Charge injection

Upon applying the voltage (taken as $t = 0$), capacitive charges are instantaneously induced on both electrodes with a surface charge density $\rho_s(t=0) = \varepsilon F$, with $\varepsilon$ being the permittivity of the dielectric and $F$ being the field intensity at the interface of the electrode. These charges can be later transferred into the sample under the force of electric field, giving arise to an injection current, the density of which can be expressed as a product of the surface charge density, $\rho_s$, and injection rate, $v_{inj}$, as:

$$J_{inj} = \rho_s v_{inj} \qquad (2)$$

For insulating polymers, intrinsic charge carriers are negligible compared to the injected charges so that initially the injection current density, $J_{inj}$, is greater than the conducting current density, $J_0$, in the bulk and electrode (normalized to the area of the electrode). From charge continuity in 1D,

$$\frac{\partial \rho_s}{\partial t} = J_0 - J_{inj} - J_{ext}^{ops} \qquad (3)$$

Here $J_{ext}^{ops}$ represents the possible extraction flux of charges of opposite polarity in the case of bipolar injection. Initially $\frac{\partial \rho_s}{\partial t}$ is negative given that the current is dominated by extrinsically injected charges and thus $\rho_s$, $J_{inj}$ decrease with time as the charges are injected into the bulk, until equilibrium is reached $\partial \rho_s / \partial t = 0$ at $t = t_0$.

As shown in Fig.1a, most of the electrons at the cathode were injected in a short time and traveled in a packet-like shape moving towards the anode. The residual charge density on the cathode at $t_0$ is much smaller than its initial value, $\rho_s^e(t_0) \ll \rho_s^e(0)$. While for holes, due to the increase in the field at the anode, it appears that $\rho_s^h(t_0) > \rho_s^h(0)$. From (2) and (3) and the fact that the charge in the bulk is predominately negative, it can be inferred that the hole injection rate in this case is much smaller than the electron injection rate. Fig.6 shows a case at which injection rates from the two electrodes are comparable (in polyethylene), so that both electrons and holes are present in the bulk in similar quantities.

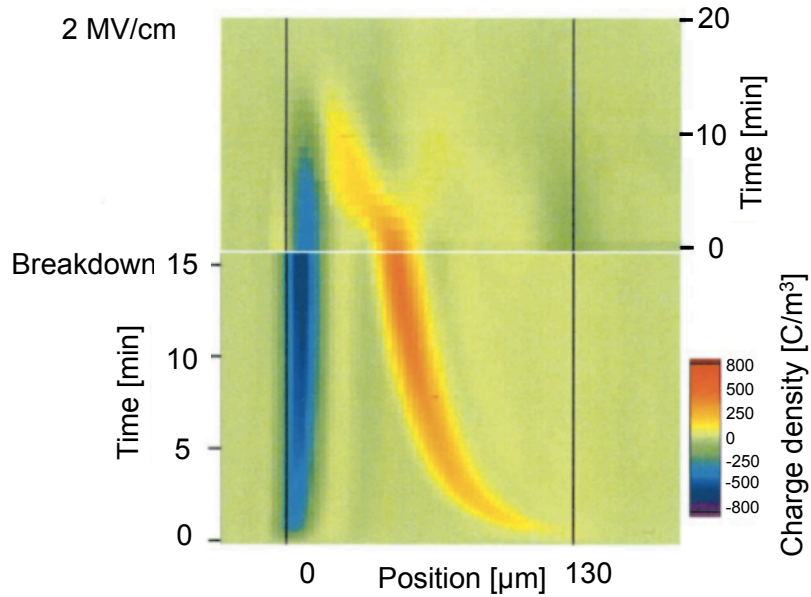

FIG. 6. The space charge distribution in a 130 μm-thick low density polyethylene (LDPE) under a nominal electrical field of 200 MV/m. The anode is made of carbon black loaded PE and the cathode is made of Al. Copyright IEEE 2005. Reprinted with permission from Ref[40].

Interestingly when the sample was stressed in an opposite voltage polarity, $\rho_s^h(t_0) \ll \rho_s^h(0)$ and positive charges predominated in the bulk, which together indicates that the hole injection rate from the SC electrode is greater than that of electrons from aluminum. By comparing results of Fig.1a and Fig.1d it can be deduced that the injection rate depends on the electrode material and that the SC electrode appear to have a larger injection rate for both electrons and holes than aluminum. However, this cannot be readily explained in terms of the difference in the injection energy barrier because barriers for hole and electron injection should add up to a constant that is equal to the energy gap of the polymer. As the injection is likely to be a tunneling process and the tunneling probability scales with $\sim \exp(-2\gamma R)$, the difference may lie in $\gamma$, which is the coupling integral between two electronic wave functions and is likely to vary at different material interfaces.

It is also worth noting that even though different electrode materials can have drastically different injection rate, the measured conductivity in this case seems to be rather insensitive to the electrode material and is likely to be limited by the bulk process. $J_{inj}$ at equilibrium ($\partial \rho_s / \partial t = 0$) is invariant to the choice of electrode (Fig.3). But the injection rate can significantly impact the charge partition between the electrode and the bulk at equilibrium, as a higher injection rate tends to leave fewer charges on the electrode and thus more in the bulk.

Here we can estimate the injection rate of electron or hole from the SC electrode. The measured $J_0$ at quasi-steady state from the dc conduction test is on the order of $0.5\,\mu\text{A}\cdot\text{m}^{-2}$ (Fig. 3). Since the charge carriers in the bulk are dominated by the ones injected from the SC electrode, it is fair to assume a small $J_{ext}^{ops}$ compared to $J_{inj}$ at SC electrode so that $J_0(t_0) \approx J_{inj}(t_0)$. Given that $\rho_s(t_0) \ll \rho_s(0)$, $v_{inj}$ is calculated to be on the order of $10^{-1}$ s$^{-1}$ as $\rho_s(t_0)/\rho_s(0)$ is found to be on the order of $10^{-2}$ (Fig.1a, Fig.1d, Fig.6). Note that this rate is for small field conditions since the field at the electrode scales linearly with $\rho_s$ and $\rho_s(t_0)$ is quite small. This rate is so large that allows half of the initial capacitive charges on the electrode to be injected within 2 seconds. It has been shown that the charge carrier is most likely to be injected to the transport level $E_t$[25] and therefore we can estimate the position of $E_t$ with respect to the Fermi level of the electrode from (1). Taking a reasonable value of $\gamma = 5$ Å$^{-1}$ and $R$ between 1~7 nm depending on the density of the localized states, the energy barrier is calculated to be in the range of 0.77 ~ 0.05 eV with the field and image charge effect specifically included[1]. This is much smaller than is predicted from the Schottky-Mott rule of several eV for wide band gap insulators. A similar large charge injection rate is also observed in polyethylene (see Fig.6)[40]. These findings indicate that the localized states can extend very deep into the energy gap of the insulator with an unnegligible concentration so that the tunneling to these states is likely to dominate the injection process. A careful reconsideration of the charge injection mechanism of the metal/insulator interface is certainly warranted and it is suggested that the localized states in insulators are of crucial importance, which although low in density might totally change the energetics governing the charge flow. The trap energy distribution and density for this case is difficult to obtain directly from the injection rate and is further investigated using a Monte Carlo simulation in later sections.

### B.   Charge recombination

Note that to exhibit the kind of bipolar charge distribution shown in Fig.1c, the recombination rate between holes and electrons must be small compared to the rate of injection. The mobility of the electron packet is on the order of $10^{-14}$ m$^2$ V$^{-1}$s$^{-1}$, which gives a recombination coefficient of $6\times10^{-19}$ m$^3$/s according to the Langevin equation[43]. It gives a lifetime as long as 60 s when the charge carrier density is in the order of $10^{20}$ m$^{-3}$. This low recombination rate is further supported by the charge distribution evolution during stressing as well as during decay after stress removal. As shown in Fig.1b, it was found that the central region of the sample changed from neutral to being positively charged and the region near the cathode changed from negatively charged to positively charged with time, implying that the initially injected electrons and holes both accumulate in these regions but electrons have a much faster decay rate than holes. This excludes recombination as the primary charge dissipation mechanism despite the spatial overlap. Otherwise the decay rate for both types of carriers should be similar and we would not see the change in charge polarity.

### C. Injected charges related ageing

The quantity of electrons in the charge packet continuously increases as it moves towards the anode and the field between the anode and the packet also increases. However, from Fig.1a it is clear that newly injected electrons do not add a long tail to the electron packet, suggesting that they traveled at a higher speed than the electron packet. And because the field behind the packet is much smaller than the region ahead of it, to fulfill this condition, the electron mobility in the region behind the packet should be much larger than that of the region ahead. This abrupt mobility change was initially explained by the trap-filling effect proposed by Matsui *et. al* in PE[40] (Fig.6): the deep traps are filled up in the region that is swept by the charge packet, so that newly injected charges hop mostly via shallow traps and can have a much larger mobility. But this explanation is not fully supported, because for effective trapping, the deep trap concentration should be at least comparable to the density of electrons in the packet, but no obvious electron accumulation in this region is ever seen[40,44].

To further investigate this phenomenon, the sample was stressed a second time after short-circuiting for 1 hour to allow the majority of traps to be emptied. The result is shown in Fig.7. It was found that during the second stress cycle, the electrons travel at a much higher speed in the region swept by the electron packet during the first cycle but then drop to the same speed as they enter the un-swept region. For even longer charge dissipation time, up to 72 hours, the electron evolution changed dramatically and significant electron accumulation against the anode was observed (Fig.7b). This indicates a greater electron mobility across the entire sample provided that the extraction rate is unchanged. To exclude the possibility that water absorption during the long charge dissipation time impacted charge mobility, another sample was retested but this time annealing was conducted by wrapping the sample within an aluminum foil heated at 70 °C under vacuum overnight. A similar phenomenon (Fig.7b) was observed. These observations confirmed that the increased electron mobility after stressing is not due to the trap-filling effect but due to some permanent changes in the material. This may be interpreted in analogy to the progressive degradation in gate dielectrics that continuous trap creation and enhanced trap-assisted tunneling cause an increase in the leakage current close to breakdown[45]. In the case of a polymer, the injected charges can induce exciton formation, catalyze bond cleavage, generate free radicals and encourage oxidations as they diffuse throughout the material[46], which is likely to create more deep traps along its path so that direct tunneling between deep traps becomes more favorable and thus may explain the increase in the apparent charge mobility. The experiment also suggests that the injected charges can still remain chemically active and are able to catalyze material degradation even after the voltage is removed.

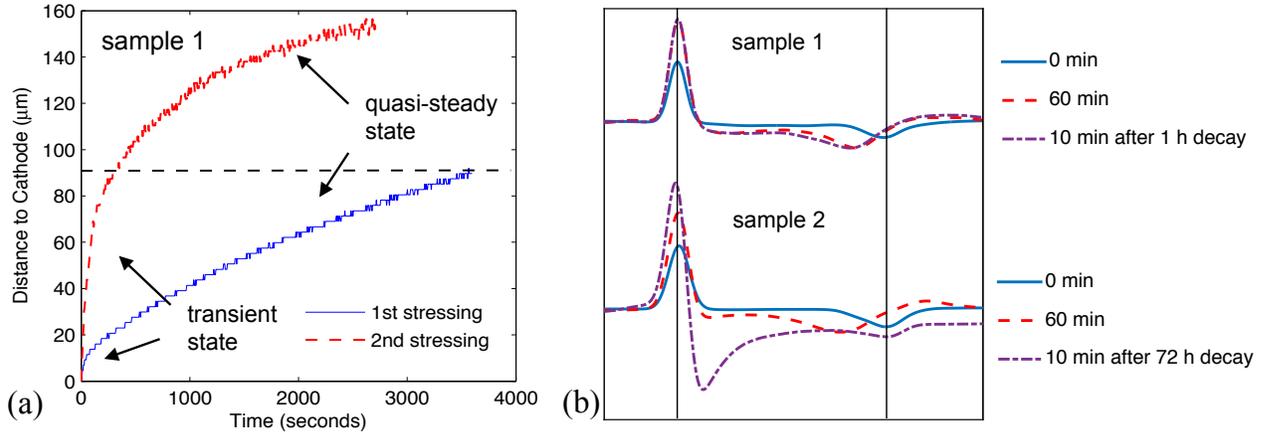

FIG. 7. (a) The position of the electron packet peak with respect to the SC cathode as a function of the stressing time. The trapped charge is allowed to decay for a hour under the short-circuit condition before the 2$^{nd}$ stressing, (b) The space charge profiles during the initial stressing and the 2$^{nd}$ stressing after different charge dissipation times.

### D. Energy distribution of traps

When the voltage is removed, the trapped charges will not disappear immediately and thus new charges are induced on both electrodes to equilibrate the electric potential on two electrodes. As shown in Fig.1b, the induced surface charges on the electrode were all positive as the trapped charges are predominately electrons. The remnant charges in the sample will decay with time by thermally activated detrapping and being extracted at the electrode. By analyzing the time dependence of the charge decay, we can probe the energy distribution of the trapped charges.

If all charges were trapped at the same depth, the detrapping rate is a constant and the total charge decay rate will just depend on the remnant charge density as

$$\frac{d\rho}{dt} = -\frac{\rho}{\tau} \qquad (4)$$

$\tau^{-1}$ describes the detrapping rate and thus characterizes the trap depth. This will give rise to an exponential decay for which $\rho = \rho_0 \exp(-t/\tau)$. If we plot $\log \rho$ with time, this should yield a linear line with a slope of $\tau^{-1}$.

If charges are trapped in states with a distribution of energies, the detrapping rate becomes heterogeneous. The charges trapped at shallower states will leave faster than those trapped at deeper states. The measured apparent charge decay rate thus will have a stronger dependence on the remnant charge density: as the remnant charge density decreases, the quasi-Fermi level of the trapped charges also decreases leading to an increasing thermal activation energy to $E_t$ and the apparent charge decay rate thus will slow down dramatically with time. The measured charge decay rate at a given time thus should be proportional to the integral of the detrapping rate over all charges,

$$\frac{d\rho}{dt} \propto - \int_{-\infty}^{E_f(t)} g(E)\exp[-(E_t - E)/kT]dE \tag{5}$$

Here $g(E)$ is the density of states and $E_f(t)$ can be approximated as the zero temperature quasi-Fermi energy of the trapped charges at time $t$ and is expected to decrease with time.

Fig.8 plots the remnant charge density as a function of time. It can be seen that the charges retained in silicone are trapped in a distribution of states, and the decay of electrons is faster than holes, implying a higher mobility. The shape of the DOS can be approximated by numerically fitting the detrapping function using multiple exponential terms with each one representing a single trapping level. The results are summarized in Fig.9. It can be seen that for electron traps, the density decreases monotonically as the traps get deeper, while for hole traps, the shape appears irregular and cannot be simply interpreted as a tail from an exponential or Gaussian distribution, likely contributed by impurities. Note that this method only allows us to probe deep traps that are filled at equilibrium with a relaxation time greater than few seconds, which covers an energy interval of ~ 200 meV. Shallower states have to be revealed by other methods with a wider range of time scales.

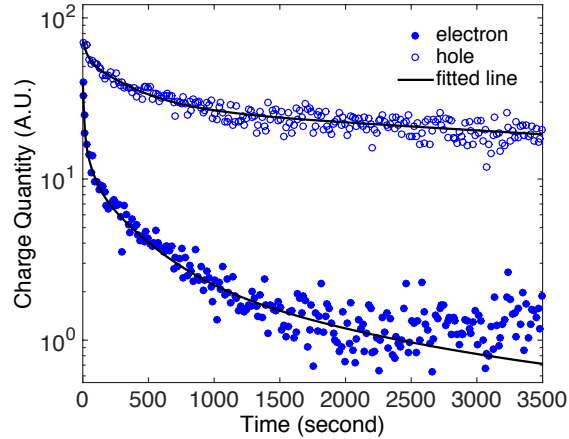

FIG. 8. The plot of remnant charge quantity in the sample as a function of time after short-circuiting two electrodes. The charge quantity is plotted in a logarithmic scale for easier comparison. The electron/hole quantity was obtained by integrating all negative/positive charges in the bulk and were fitted by multiple exponetial decay terms. The hole data is obtained from the sample stressed by a positive polarity, since in which case most holes are not shadowed by electrons.

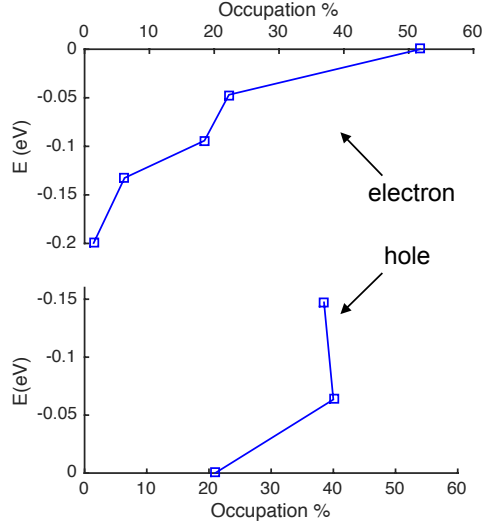

FIG. 9. The occupation of the trapped charges in different trap depth obtained from the decay rate fitting. The shape of the curve reflects the DOS of electron and hole trap states. The energy reference zero is set to be the highest measured occupied trap states of electron and hole respectively.

### E. Transient and dispersive transport

As shown in Fig.2d, the transport velocity of the electron packet was found to decrease with time at short times after the voltage application in a power-law fashion $v = t^{-(1-n)}$. $n$ was found to be close to zero for the initial 20 seconds and then gradually increased to close to one when approaching quasi-steady state. A similar power-law decay was also found in the measured transient current in the external circuit, as shown in Fig.3. The slope in the initial few minutes closely matches that observed in PEA, indicating a strong correlation. Similar transient hole transport is also exhibited in PE as shown in Fig.6.

It was suggested that the energy relaxation of charge carriers into a distribution of localized states could lead to a transient current[47]. This mechanism is first proposed to explain the power-law decaying photocurrent in amorphous semiconductors and has been well analyzed[23,48]. It was argued that at short times, only carriers at shallow states have a typical detrapping rate greater than the reciprocal of the observation time $1/t$ and thus can be treated as mobile. With increasing time, charge carriers continuously drop to deeper energy states and the rate to "detrap" should scale with the reciprocal of the observation time as $v_{typ} = 1/t$. The exponent of the power-law term is related to the occupation number of the carriers at the transport level and is determined by the shape of DOS. For dark current in dielectric polymers, the idea also applies if charge carriers are initially injected to shallower states and later trapped at deeper states as they traverse the material.

Deriving an analytical solution to capture this effect can be difficult because the local concentration of carriers varies with time and thus numerical methods are usually employed. Due to the stochastic nature of trapping and detrapping, here we performed a 3D Monte Carlo (MC) simulation to investigate the transient electron transport under the MH model. A Gaussian DOS $g(E) = \frac{1}{\sigma\sqrt{2\pi}} \exp\left(-\frac{E^2}{2\sigma^2}\right)$ is used with $\sigma$ = 0.224 eV and the band tail below -1.0 eV (reference zero set to be the mobility edge in the conduction band) is modified to match the experimentally measured electron trap distribution shown in Fig.9. The hopping rate is determined by the Abraham-Miller formula with $\gamma^{-1}$ = 5 Å at a field of 1 MV/m. The injection of electrons is assumed to be instantaneous and the Columbic repulsion is specifically considered. The details of the simulation can be found in the supplemental information[42].

Fig.10 shows that such set of parameters can give a reasonable match to the experiment results and successfully recover the transient transport regime. Fig.11 plots the spatial charge distribution as a function of time and clearly reveals a dispersive transport, in line with the observation from PEA measurements.

The energy relaxation into a local minimum was found to be a rather quick process (< 1 s), and the transient charge transport is critically determined by the relative concentrations of charge carriers and deep trap states at the band tails. In the abovementioned condition, the concentration of states with energy deeper than -1 eV is as low as $8 \times 10^{21}$ m$^{-3}$, corresponding to an average distance of 50 nm. The transient transport arises because the initial carrier density (with an average planar distance of 25 nm) is higher than the deep trap concentration so that carriers cannot all be immediately trapped at these deep states due to columbic repulsion or Pauli exclusion. In such condition, the quasi-Fermi energy of the injected carriers is initially held at a higher level and then gradually decreases with time as the carrier density decreases due to dispersive transport. This gives rise to transient transport for which the collective mobility decreases until a quasi-steady state is reached when the quasi-Fermi energy essentially drops down to the equilibrium minimum. And as shown in Fig.12, the transient traveling distance shows a negative dependence on the deep trap concentration and this is because a larger trap concentration provides more efficient trapping and also brings the initial quasi-Fermi energy closer to the equilibrium minimum.

Dispersive transport is caused by non-uniform trapping and is found to also depend on the trap concentration. Fig.13 shows that the standard deviation of the traveling distance of the electron at a given time increases with the corresponding median value, as a signature of dispersive transport. And more interestingly, increasing or decreasing the trap density tends to decrease the extent of dispersion, and the maximum dispersion seems to occur at a particular trap density. It can be argued that the maximum occurs at the trap density comparable to the charge carrier density since either increasing or decreasing the trap density tends to smooth out the variation of the trapping time (either all being trapped at deep traps or not being trapped at all).

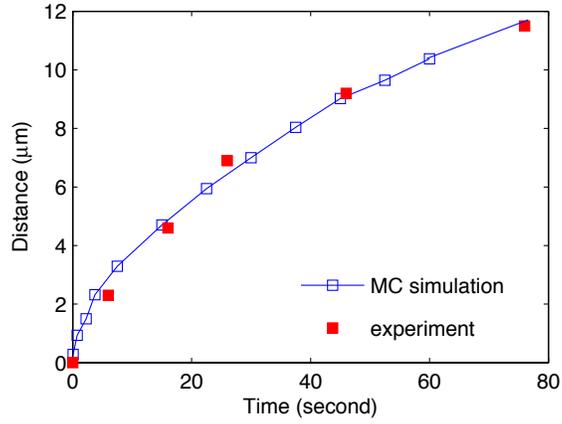

FIG. 10. The traveling distance of electron packet with respect to cathode as a function of time from both MC simulation and PEA measurement at 10 MV/mm. The data from MC simulation is taken from the median value over 320 electrons.

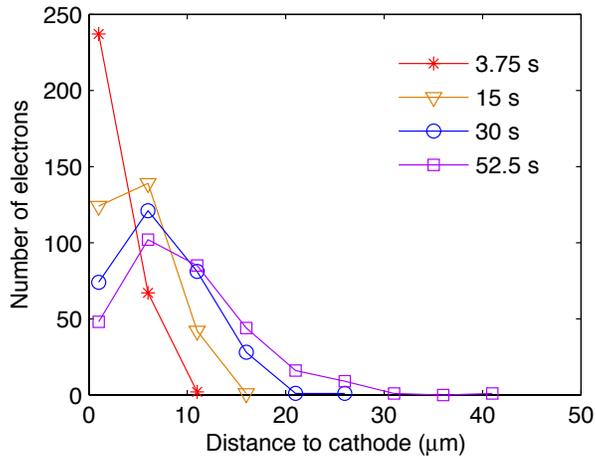

FIG. 11. The spatial electron distribution at different time obtained from MC simulation.

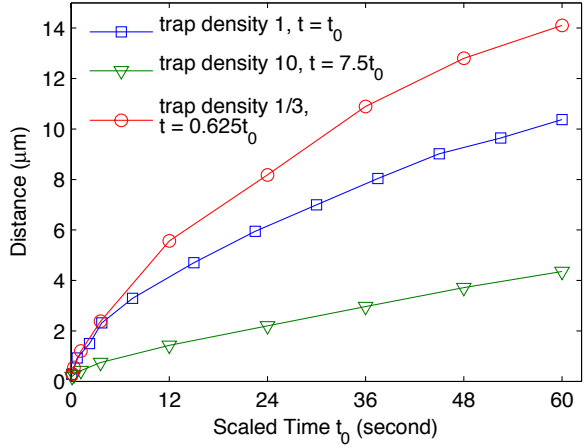

FIG. 12 The traveling distance as a funciton of time at different deep trap densities (normalized with respect to $8\times10^{21}$ m$^{-3}$). The time is rescaled for each simulation for better comparision.

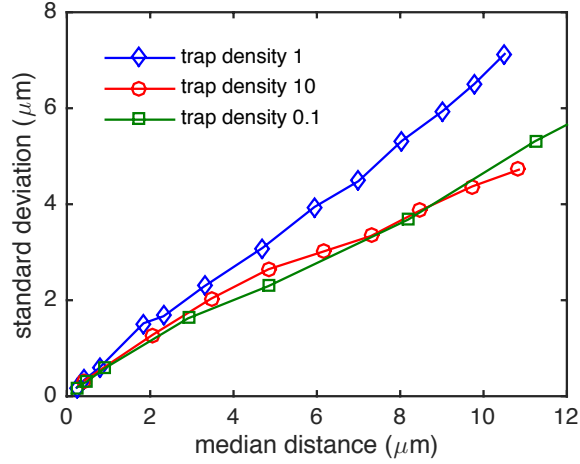

FIG. 13 The standard deviation of the traveling distance as a funciton of the median value for different deep trap densities (normalized with respect to $8\times10^{21}$ m$^{-3}$).

### F. Fast carriers and negative field dependent mobility

As shown in Fig.4, both the transient and quasi-steady state dc current in the external circuit is found to increase almost linearly with the field. This rules out the injection current as the dominate factor in the measured transient current, as the injection rate is expected to increase exponentially with the field. Also, to maintain a linear *I-V* relationship, $\rho_s(t_0) \propto \ln(F)$. So that the electron packet formation is maintained with an approximate equivalent planar charge density of $\varepsilon V/L$ with *V* being the applied voltage and *L* being the distance between the charge packet and anode, just as shown in Fig.2.

From Maxwell's equation it can be derived that

$$\nabla \cdot (\nabla \times \mathbf{H}) = \nabla \cdot \left( \mathbf{J} + \frac{\partial \mathbf{D}}{\partial t} \right) = 0 \tag{6}$$

where $\mathbf{J}$ is the free charge current and $\mathbf{D}$ is the electric displacement. In a 1D case, (6) basically states that the sum of free charge current density and displacement current density is a constant everywhere in the circuit. If the evaluation point is taken in the region between the electron packet and the anode, $\partial D/\partial t$ is non-zero due to the movement of the electron packet. And its value can be approximated as $J_p \Delta x / L$, if the electron packet is treated as a sheet of charges with a thickness of $\Delta x$ and the $J_p$ is the current density within the charge sheet. So $J_0$ can be written as

$$J_0 = J_c + J_p \frac{\Delta x}{L} \tag{7}$$

with $J_c$ as the conducting current due to free charge carriers in the evaluated region. The second term of (7) can be directly evaluated from PEA and appears to be rather invariant with the field given that the charge density and velocity shows an opposite trend. By comparing with the measured $J_0$, it was found that $J_c$ constitutes more than 85% of $J_0$ and the ratio further increases at higher field. As a result, $J_p/J_c$ decreases from 0.5 to 0.08 as the field increases from 10 MV/m to 50 MV/m. PEA reveals that the charge carriers contributing to $J_c$ have a much lower density ($\rho_{fast}$) compared to the electron packet so that these charges must have traveled at a much higher speed ($u_{fast}$) than $u_p$ in order to form a large current. If $\rho_{fast}$ is 1 % of the charge density of the electron packet, then $u_{fast}/u_p \sim 10^3$, corresponds to an activation energy of 0.18 eV from Boltzmann statistics. More interestingly, the velocities of two types of charge carriers show opposite trends: few fast charge carriers shows an increasing velocity with the field in contrast to the decreasing velocity of the slow moving electron packet that consists of the majority of the injected electrons.

The nature and transport mechanism of these fast carriers, however, is not well understood. The evidence of their existence was initially found in the time-of-flight mobility measurement of electron beam excited electron/hole in late 1970s in various polymers[49,50]. Recent experiments utilizing an ultrafast PEA acquisition technique further revealed that these charges exhibit a soliton-like transport with a pulsed generation[50,51]. The measured mobilities from both techniques fall into the range of $10^{-11}$ to $10^{-9}$ m$^2$/Vs depending on the polymer type and do not exhibit negative field dependence. The transport was found to be thermally activated and was proposed to depend on the polymer side group relaxations[52]. It is difficult to imagine that the electronic charges in dielectrics can have two separate transport mechanisms that are entirely isolated from each other. On the contrary, our results show that the transport of slow and fast charges are actually closely related as they both exhibit a similar transient behavior and we have found that adding nanoparticles can simultaneously decrease the current of both [2nd paper] (This manuscript is submitted in a series with another one and this is the cross-reference to the 2nd one).

Therefore we speculate that these fast carriers are nothing special but are those that are "lucky" enough to escape most deep traps as they transverse the sample. They contribute to the highly asymmetrical long tail at the front of charge distribution as hinted in Fig.11. It should be noted that the local concentration of un-trapped and trapped charges can deviate significantly from the thermal equilibrium, especially for short traverse distances, small trap density as well as large spatial heterogeneity of trap distribution that may arise intrinsically or extrinsically from defects and impurities. And this explains why the fast and slow charges can have drastically different spatial distributions.

The negative field dependent transport velocity of the charge packet is not a unique feature for silicone but has also been found in other polymers like PE[40,53] and PMMA[54]. The mechanism however still remains unclear and cannot be readily explained by previously cited transport models since they all predict an exponential increase of mobility with field $F$ or $F^{1/2}$ [55]. Normally we would expect the velocity to increase with the field because (1) the injected carrier density increases at high field so that more carriers can travel at shallower states; (2) the energy barrier is further reduced at higher fields. Chen and Zhao[53] related this anomalous phenomenon in PE to the Gunn effect found in the band transport of some semiconductors that high field can push carriers to higher energy states characterized with a larger effective mass and thus a smaller mobility. This theory however cannot explain why few carriers can still travel at an increasing speed with the field and it also gives unrealistic predications like a shrinking rather than spreading distribution of the charge packet with time[56].

This phenomenon can also be addressed within the hopping transport model for specific disordered systems. Studies suggested that the anomalous decrease of velocity especially at high fields is an inherent feature for carriers hopping in disordered systems characterized with large positional disorder in addition to the energetic disorder[57,58]. The positional disorder causes a variation in the geometrical distribution of localized states which creates spaces that are void of low energy states. Normally the charge carrier can find a more favorable detour to avoid tunneling through large barriers or long distance, but increasing the field will diminish the probability for jumps in the direction perpendicular or against the field. Higher field imposes a larger directionality on carrier hopping and thus blocks the detour path. Both analytical theories and Monte Carlo simulations have been developed to explain and validate the phenomenon in the hopping regime[58,59]. This explanation correctly predicts the increasing discrepancy for velocities of the fast and slow moving carriers because "lucky" carriers that have avoided geometrical traps are not subject to this effect and will travel with an increasing speed with the field.

### G. Effect of SiO$_2$ filler

For comparison purpose, the space charge evolution for a silicone matrix that is free of SiO$_2$ fillers is plotted in Fig.14. The transient regime traveling distance was found to be much longer, and the electron packet travels at a speed ~5 times larger than the SiO$_2$ reinforced silicone. The dispersion in electron mobility also appears to be more pronounced as the electron packet quickly spreads out with time. Based on previous discussions, we can deduce that the unfilled silicone has much fewer deep traps. But since the $SiO_2$ filler loading is as high as 40 wt%, it is uncertain whether these deep traps are due to the extrinsic energy states on the $SiO_2$ surface or are intrinsic to the polymer but induced from the morphology change caused by the $SiO_2$. The effect of nanoparticles will be further pursued in our next paper.

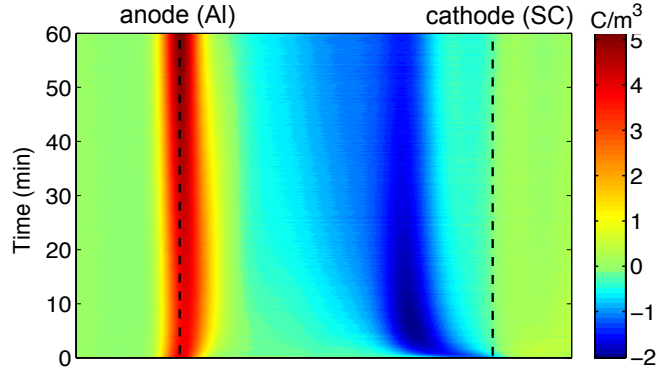

FIG. 14. The space charge profile for unfilled silicone matrix at 10 MV/m.

## V. CONCLUDING REMARKS

Rather than being regarded as a perfect wide band gap semiconductor, polymer dielectrics contain many localized states in a distribution of energy due to either morphological disorder, polarization, chemical defects or impurities. Systems with similar electronic structures can also be found in amorphous organic and inorganic semiconductors and the developed hopping transport models may be well applied to dielectric polymers. Several key findings are presented in this work and are explained within the multiple-hopping model.

1. The localized states can extend very deep into the forbidden gap and thus render the Schottky junction model inaccurate to describe the charge transfer rate at the polymer/metal interface given that tunneling directly to these states is likely to dominate the injection process.

2. A transient hopping transport can occur if the initial density of injected carriers is greater than that of deep traps. The density and energy of these deep traps critically determine the effective mobility as well as the persistence time and distance of the transient transport.

3. Not only the energy distribution but also the geometrical distribution of the localized states are important to charge transport. And a large disorder in geometrical distribution is likely to be responsible for the observed negative field dependent mobility of the charge packet.

4. Because of non-uniform trapping, the charge transport is highly dispersive and a few carriers that have avoided trapping can travel at a much higher speed and dominate the conducting current at high field.

5. The trap concentration is not static but subject to change as a consequence of electrical ageing and may account for the observed raised charge mobility after stressing due to enhanced trap-assisted tunneling.

It can be seen that the charge transport properties in dielectric polymers are largely determined by the localized states especially those with low energies that are characterized as deep traps. There have been both experimental and computational efforts in tracing the origin of these traps and quantifying their density and energies, including traps due to the polaron effect[60], structural disorder and cavities[61], and chemical defects and impurities[15,16,62,63]. But a comprehensive picture of the relationship between the trapping parameters and the polymer chemistry, processing condition and the electrical ageing is still missing, posing difficulties for material design. In addition, it has been shown that the addition of specific nanoparticles and organic fillers could help with electric endurance by introducing extrinsic traps and their effect will be further pursued in the following paper.

## ACKNOWLEDGMENTS

The authors wish to acknowledge the financial support from the Office of Naval Research and express thanks to Prof. Keith Nelson and Prof. Toh-Ming Lu, Rensselaer Polytechnic Institute for many useful discussions.